\documentclass[aps,pra,reprint,superscriptaddress,10pt,longbibliography]{revtex4-1}

\usepackage{graphicx}
\usepackage{amsmath}
\usepackage{amssymb}
\usepackage{amsfonts}
\usepackage{bm}
\usepackage{bbm}
\usepackage[mathcal]{euscript}

\usepackage{multirow}

\usepackage{pythonhighlight}

\usepackage{color}

\newcommand{\etal}{\textit{et al}.}

\begin{document}
\title{Parametrization of the Tkatchenko-Scheffler dispersion correction scheme for popular exchange-correlation density functionals:
effect on the description of liquid water}

\author{Miguel~A. Caro}
\email{mcaroba@gmail.com}
\affiliation{Department of Electrical Engineering and Automation,
Aalto University, Espoo, Finland}
\affiliation{COMP Centre of Excellence in Computational Nanoscience,
Department of Applied Physics, Aalto University, Espoo, Finland}

\newcommand{\eq}[1]{Eq.~(\ref{#1})}
\newcommand{\fig}[1]{Fig.~\ref{#1}}

\begin{abstract}
We present a list of optimized damping range parameters $s_R$ to be used with the Tkatchenko-Scheffler
van der Waals dispersion-correction scheme [Phys. Rev. Lett. 102, 073005 (2009)]. The optimal
$s_R$ are obtained for seven popular generalized-gradient approximation exchange-correlation density
functionals: PBE, RPBE, revPBE, PBEsol, BLYP, AM05 and PW91. The optimization is carried out in the
standard way by minimizing the mean absolute error of the S22 test set, where the reference interaction
energies are taken from coupled-cluster calculations. With the optimized range parameters, we assess
the impact of van der Waals corrections on the ability of these functionals to accurately describe structural
and thermodynamic properties of liquid water: radial distribution functions, self-diffusion coefficients
and standard molar entropies.
\end{abstract}

\date{\today}

\maketitle

\section{Introduction}

The Tkatchenko-Scheffler (TS) method has emerged in recent years as one of the most popular dispersion-correction schemes in density
functional theory (DFT) calculations~\cite{tkatchenko_2009}. The reason is that TS is a post-processing scheme, and can therefore
be applied on top of DFT electron densities without significantly adding to the overall execution time of the different algorithms,
where by far the most expensive part corresponds to the self-consistent calculation of the Kohn-Sham orbitals. It was shown by Marom
\etal~\cite{marom_2011} that TS performs better than similar approaches for a wide variety of weekly interacting molecular systems.

The TS approach involves the optimization of an adjustable parameter, $s_R$, which determines the range at which the dispersion
interactions begin to become important. The value of $s_R$ depends on the intrinsic ability of the underlying exchange-correlation (XC)
density functional to correctly describe van der Waals interactions. An optimal value of $s_R$ is therefore functional-dependent.
Tkatchenko and Scheffler originally optimized $s_R$~\cite{tkatchenko_2009} for the PBE functional~\cite{perdew_1996} by minimizing
the error in the interaction energies predicted with the PBE+TS approach compared with highly accurate coupled-cluster calculations [CCSD(T)] of
the S22 test set by Jure\v{c}ka \etal~\cite{jurecka_2007}. They obtained $s_R = 0.94$. Later on, Marom~{\etal} optimized $s_R$ for a number
of hybrids and meta-generalized-gradient approximation (meta-GGA) XC functionals~\cite{marom_2011}. Agrawal~{\etal} carried
out follow-up work on the performance of TS corrections for hybrid functionals, including the role of the range-separation screening
parameter, finding it to have little correlation with the optimization of $s_R$~\cite{agrawal_2013}. However, optimal $s_R$ have not yet been
estimated for many popular GGA functionals, which are routinely used preferentially over more advanced functionals because
of computational advantages. $s_R$ is not transferable across XC functionals, and using for instance the PBE value with a different GGA
may lead to unacceptable errors in the calculated dispersion corrections.

\section{Optimization of $s_R$ for several XC functionals}

In this paper we optimize $s_R$ for some of the GGAs most widely used across the computational chemistry and physics communities:
PBE (for benchmark with previous results),
RPBE~\cite{hammer_1999}, revPBE~\cite{zhang_1998}, PBEsol~\cite{perdew_2008}, BLYP~\cite{becke_1988,lee_1988,gill_1992},
AM05~\cite{armiento_2005} and PW91~\cite{perdew_1992}.
We use the GPAW DFT suite~\cite{enkovaara_2010} in conjunction with the Atomic Simulation Environment (ASE)~\cite{larsen_2017}, which offer
an extremely flexible Python-based environment to carry out the present calculations (a sample Python script is presented at the end of this
paper that allows to optimize $s_R$ for any XC functional available from the LibXC library~\cite{marques_2012} and compatible with GPAW).
We used the GPAW grid mode with 0.18~{\AA} spacing and PAW potentials~\cite{bloechl_1994,kresse_1999}. We generated the PAW setups for the
PBEsol, BLYP, AM05 and PW91 functionals using the GPAW setup generation tool. The DFT calculations were carried out in fixed boundary conditions
within orthogonal boxes where at least 4~{\AA} of vacuum was allowed between the atoms and the box boundaries.

\begin{figure}[t]
\includegraphics{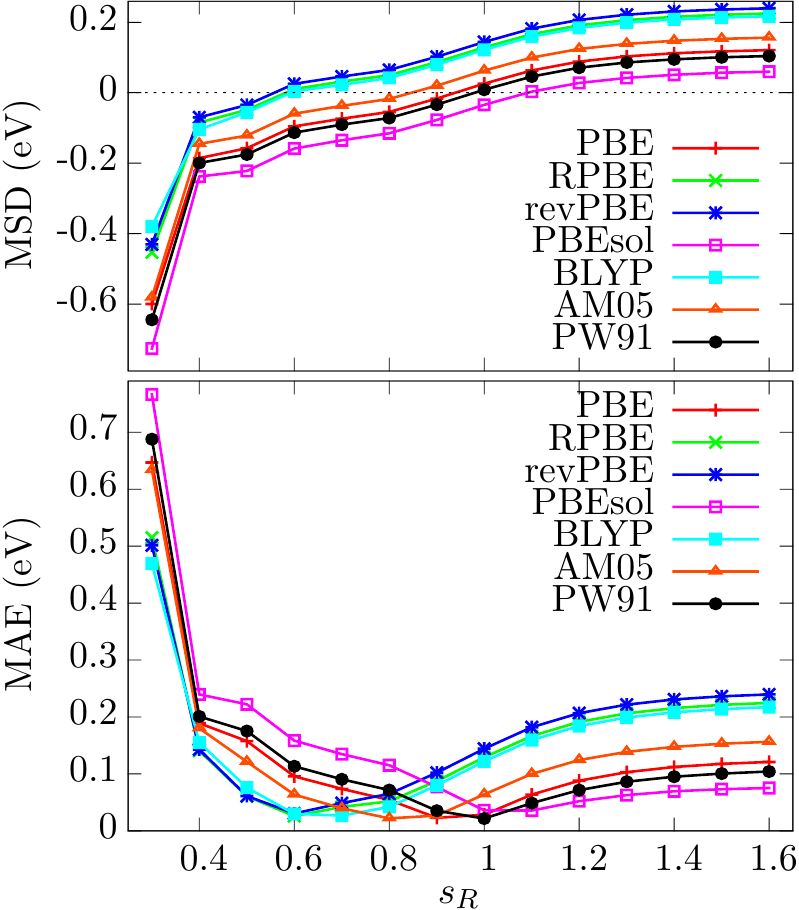}
\caption{Mean signed difference (MSD) and mean absolute error (MAE) of the TS method applied to different GGAs over a
wide range of $s_R$ values.}
\label{01}
\end{figure}

\begin{figure}[t]
\includegraphics{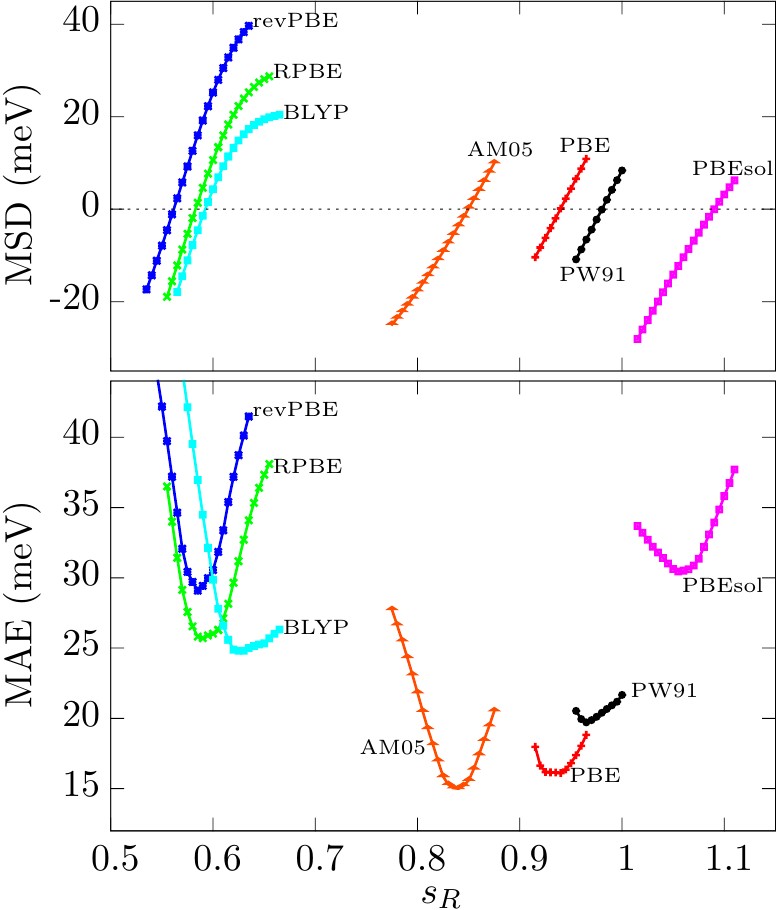}
\caption{Fine detail of \fig{01} in the region where the errors are minimized.}
\label{02}
\end{figure}

\begin{figure*}[t]
\includegraphics{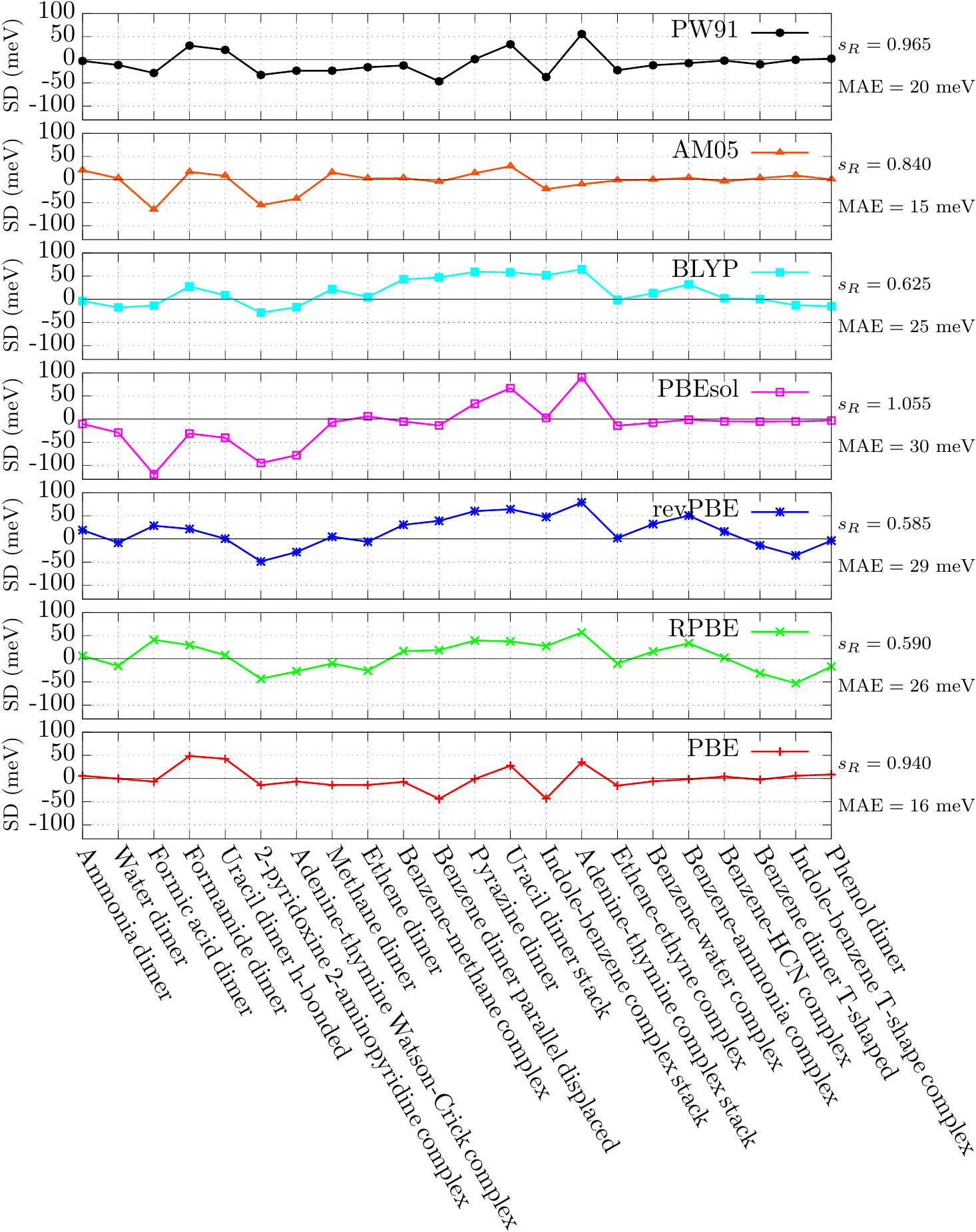}
\caption{Performance of the DFT+TS scheme for different GGA XC functionals, evaluated at each functional's optimal
$s_R$ value for all and each of the molecular systems present in the S22 set.}
\label{04}
\end{figure*}

\begin{table}[t]
\caption{Optimal $s_R$ fitted for each XC functional (resolved to nearest 0.005) and MAE calculated for the S22
test set. The calculations for each functional were run at the optimal value of $s_R$ for that functional.}
\begin{ruledtabular}
\begin{tabular}{l c c}
XC functional & $s_R$ & MAE (meV)
\\
\hline
PBE & 0.940 & 16
\\
RPBE & 0.590 & 26
\\
revPBE & 0.585 & 29
\\
PBEsol & 1.055 & 30
\\
BLYP & 0.625 & 25
\\
AM05 & 0.840 & 15
\\
PW91 & 0.965 & 20
\end{tabular}
\end{ruledtabular}
\label{03}
\end{table}

The interaction
energies are defined as the difference between the total energy of the interacting system (1+2) and the sum of the energies of
each of the two isolated molecules (1 and 2):
\begin{align}
E_i^\text{int} = E_i^{1+2} - E_i^1 - E_i^2,
\end{align}
where $i$ labels each of the 22 molecular systems in the S22 set. The mean absolute error (MAE) and mean signed difference (MSD)
between the interaction energies calculated with GPAW
and those from the CCSD(T) S22 test set computed for each XC functional over a wide range of $s_R$ values are shown in \fig{01}.
One can observe the same overall trends regardless of the GGA used: the error increases rapidly as the value of $s_R$ is reduced
and slowly as it is increased. It can also be observed how the MAE for each functional has its minimum at a different position,
emphasizing the fact that $s_R$ values are not transferable across XC functionals and must be carefully optimized on an
individual basis.

Figure~\ref{02} shows a detailed view of MSD and MAE in the regions where they are zero and minimum, respectively. Table~\ref{03}
summarizes the optimal $s_R$ parameter obtained for each XC functional studied and the MAE computed at said value of $s_R$. We obtain
the same value for PBE as has been previously reported~\cite{tkatchenko_2009,marom_2011}. As discussed by Marom \etal~\cite{marom_2011},
a low optimal $s_R$ indicates that the underlying XC functional does a poor job at handling dispersion interactions on its own.
From \fig{02} and Table~\ref{03} we observe
how revPBE, RPBE and BLYP cluster around the low $s_R$ region, while AM05, PBE and PW91 cluster at intermediate values and PBEsol lies
slightly beyond them. We also observe that AM05 and PBE see their MAE reduced to a very small value ($\sim 15$~meV) with the inclusion of
TS dispersion corrections, while revPBE and PBEsol retain larger errors of up to 30~meV even with the optimal $s_R$.

To get deeper insight into the different contributions to the overall errors, \fig{04} shows the signed errors (SD) for each molecule
in the S22 set and each functional, where the TS corrections are computed at the optimal $s_R$ value for each XC functional. Note
from the figure that the description of dispersion is not at all homogeneous across molecules. For instance, PBEsol+TS, which performs
worst overall, does a very good job at describing many of the benzene-containing systems. We also note how several functionals
give almost identical results to each other, for instance PW91 and PBE show almost identical curves. The same is true for RPBE and
revPBE.

Based on these observations, the possibility to re-optimize $s_R$ for a very specific problem by fitting to a more suitable test
set remains viable. For instance, if one is interested in achieving an accurate description of a protein, $s_R$ could be optimized
to describe the dispersion interactions between a set containing different aminoacids and small organic molecules.
The Python script in the Appendix automatizes the calculation of dispersion errors
presented in this paper, and can be used to optimize $s_R$ for a different XC functional or test set (note that one needs to generate
the corresponding PAW setups needed by GPAW beforehand if they are not installed by default).

\section{Application of dispersion corrections to describe liquid water}

\begin{figure*}[p]
\includegraphics{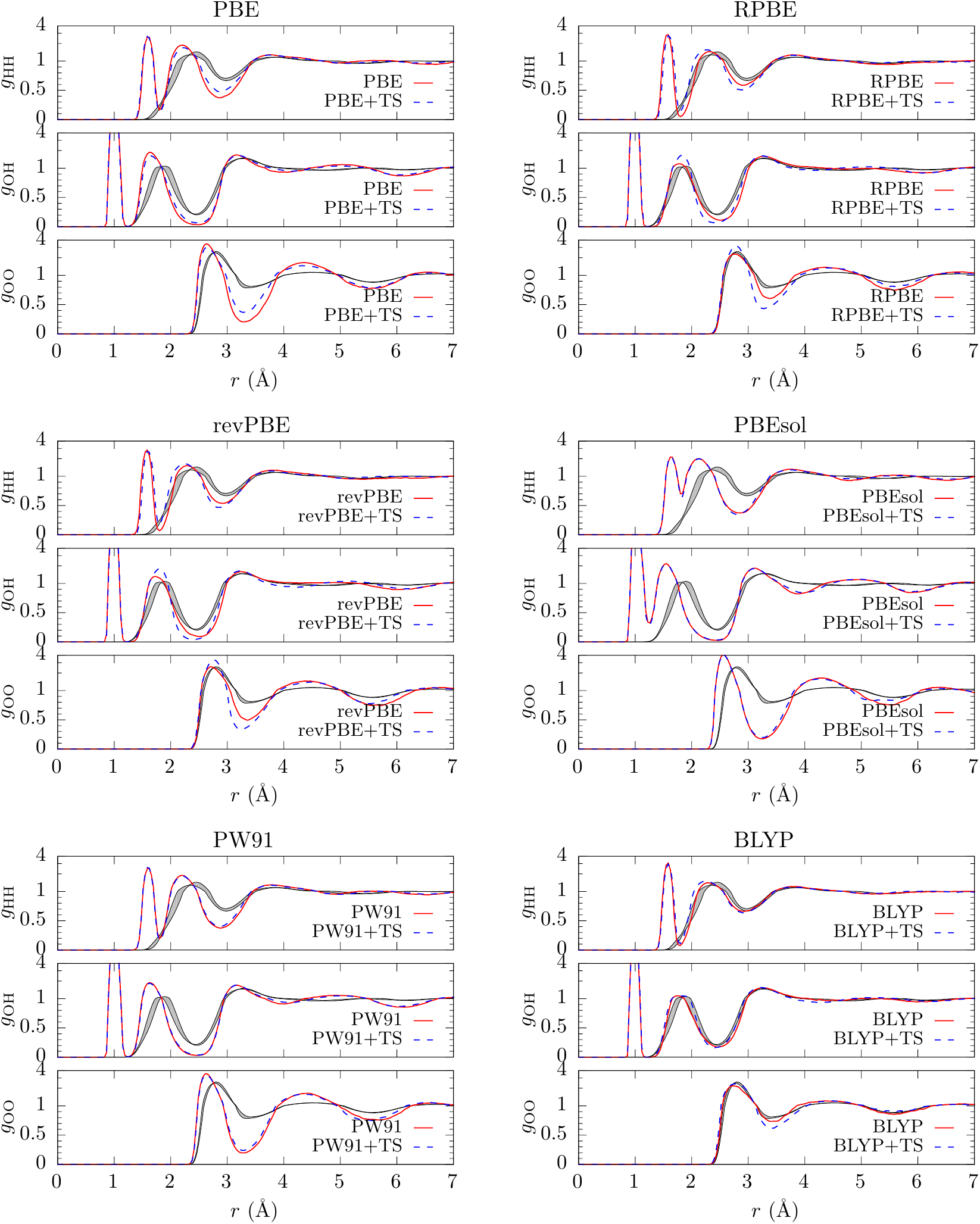}
\caption{Radial distribution functions calculated for the different GGAs studied in this paper and their dispersion-corrected counterparts ($T = 300$~K
and $\rho = 1$~g/cm$^3$).
The shaded curves indicate experimental results at $T = 298$~K (including error margins) from Soper~\cite{soper_2013}. The peaks corresponding to
intramolecular bonds are not shown by the experimental curves. The vertical axes are linear from 0 to 1 and logarithmic from 1 to 4.
Numerical values for the positions of the different maxima and minima are given in Table~\ref{07}.}
\label{06}
\end{figure*}

To test the possible improvement over the underlying functionals brought about by the inclusion of dispersion corrections,
we looked at molecular dynamics (MD) simulations of liquid water. A more exhaustive assessment of the ability of DFT to
correctly describe the properties of water is available from the recent excellent review by Gillan, Alf\`e and Michaelides~\cite{gillan_2016};
here we are more concerned with identifying systematic improvements introduced by TS dispersion corrections. A detailed study
along the same lines was conducted for the BLYP, PBE and revPBE functionals~\cite{lin_2012}, although hydrogen nuclei were replaced by deuterium
nuclei in that study. This has the observed effect of decreasing the self-diffusion constant of the liquid.
In this study, the performance of the different GGAs for predicting structural and
thermodynamic properties of water at ambient conditions (300~K) and experimental density (1~g/cm$^3$) was studied with and without
TS corrections. The properties that we looked at are: self-diffusion coefficient, radial distribution functions and standard molar
entropy. We used periodic supercells containing 100 water molecules, pre-equilibrated with a classical potential and the Gromacs
code~\cite{abraham_2015}. We used the SPCE water model~\cite{berendsen_1987}
(allowing for vibrations of the O-H intramolecular bonds) with the OPLS force field~\cite{jorgensen_2005}. After this,
35~ps of \textit{ab initio} MD (AIMD) followed. The AIMD part was performed with the plane-wave based DFT code VASP~\cite{kresse_1996}.
Since PAW potentials for VASP are only supplied for the PBE and LDA functionals, we carried out the AIMD using PBE PAW potentials for
all the GGAs studied here. We used 0.5~fs as time step to correctly resolve hydrogen vibrations, and set the kinetic energy cutoff
for the plane-wave basis set to 300~eV. The first 15~ps of dynamics for each simulation were discarded and the last 20~ps were analyzed
with our own implementation~\cite{caro_2016,caro_2017b} of the 2PT method~\cite{berens_1983,lin_2003,lin_2010}, DoSPT~\footnote{{http://dospt.org}}.
Unfortunately, we could not compute the AM05 values due to convergence problems in the MD runs. The values for all the other GGAs
tested and reference experimental and classical MD values are given in Table~\ref{05}. Radial distribution functions are shown in \fig{06} and
detailed information regarding RDF peak positions is given in Table~\ref{07}.

\begin{table}[t]
\caption{Standard molar entropies $S_0$ and self-diffusion coefficients $D$ calculated for different GGAs with and without
TS dispersion corrections at $T=300$~K. Experimental and classical MD results are shown for comparison.}
\begin{ruledtabular}
\begin{tabular}{l c c}
XC functional & $D$ ($10^{-9}$ m$^2$/s) & $S_0$ (J/mol/K)
\\
\hline
PBE & 0.57 & 43.2
\\
PBE+TS & 1.09 & 47.8
\\
RPBE & 1.61 & 54.4
\\
RPBE+TS & 1.09 & 49.1
\\
revPBE & 1.28 & 51.5
\\
revPBE+TS & 0.66 & 45.5
\\
PBEsol & 0.40 & 39.1
\\
PBEsol+TS & 0.49 & 40.8
\\
BLYP & 2.16 & 58.0
\\
BLYP+TS & 1.59 & 54.7
\\
PW91 & 0.64 & 42.8
\\
PW91+TS & 0.92 & 45.0
\\
\hline
OPLS+SPCE (rigid) & 2.72 & 58.8
\\
OPLS+SPCE (flexible) & 1.54 & 53.3
\\
\hline
Experiment & 2.41\footnotemark[1] & 69.92\footnotemark[2]
\end{tabular}
\end{ruledtabular}
\footnotetext[1]{From Ref.~\cite{holz_2000}.}
\footnotetext[2]{From Ref.~\cite{nist_69} at $T=298$~K.}
\label{05}
\end{table}

In Ref.~\cite{gillan_2016}, uncorrected RPBE was observed to offer the best description of liquid water among the
tested GGAs. Here, we observe the same behavior for RPBE but obtain best results in all the categories, including entropy,
which was not surveyed in Ref.~\cite{gillan_2016}, for BLYP. The main difference between the sources quoted in Ref.~\cite{gillan_2016}
and this work is that all MD here were carried out with regular hydrogen nuclei (protium) rather than deuterium. We also
ran all the simulations at the experimental density. Heavy water has a lower self-diffusion constant than regular water,
and a quick test that we ran with the classical potential we employed for pre-equilibration (SPCE+OPLS) showed that
the self-diffusion constant decreases by 40\% when using deuterium nuclei instead of protium nuclei.

The revPBE functional shows very similar performance to the related RPBE functional. Surprisingly, the inclusion of TS dispersion corrections
worsens the agreement with experiment for both of those functionals and BLYP, while it improves it for the others. PBEsol emerges
as the absolute loser in this comparison. The bad performance of PBEsol does not only affect the quantitative description
of the properties of liquid water, but it also gives a qualitatively bad behavior of the water molecules during the MD. We
observed significant spontaneous hydroxyl and hydronium ion formation with PBEsol: 4\% of the oxygens were observed to be part of either
a hydroxyl or a hydronium group at any time of the dynamics. This can be graphically observed as the lack
of a clear minimum in the O-H radial distribution function in \fig{06}. Therefore, we conclude that the PBEsol functional should be avoided
for any simulation that involves water molecules, such as simulations of solid/water interfaces, in order to prevent unphysical results.

\begin{table*}[t]
\caption{Position of the first two maxima and first two minima of the RFDs of liquid water depicted in \fig{06}.}
\begin{ruledtabular}
\begin{tabular}{l c c c c c c c c}
XC functional & $r^{(1)}_\text{max}$ (\AA) & $g^{(1)}_\text{max}$ & $r^{(1)}_\text{min}$ (\AA) & $g^{(1)}_\text{min}$
& $r^{(2)}_\text{max}$ (\AA) & $g^{(2)}_\text{max}$ & $r^{(2)}_\text{min}$ (\AA) & $g^{(2)}_\text{min}$
\\
\hline
 & \multicolumn{8}{c}{H-H RDF}
\\
PBE & 1.59 & 2.59 & 1.80 & 0.17 & 2.20 & 1.88 & 2.88 & 0.37 
\\
PBE+TS & 1.59 & 2.64 & 1.80 & 0.14 & 2.23 & 1.69 & 2.85 & 0.48 
\\
RPBE & 1.59 & 2.83 & 1.80 & 0.05 & 2.31 & 1.45 & 2.92 & 0.58 
\\
RPBE+TS & 1.59 & 2.72 & 1.80 & 0.14 & 2.27 & 1.56 & 2.92 & 0.51 
\\
revPBE & 1.59 & 2.80 & 1.80 & 0.07 & 2.27 & 1.52 & 2.92 & 0.54 
\\
revPBE+TS & 1.59 & 2.66 & 1.80 & 0.17 & 2.23 & 1.64 & 2.85 & 0.47 
\\
PBEsol & 1.62 & 2.14 & 1.80 & 0.70 & 2.13 & 2.00 & 2.85 & 0.37 
\\
PBEsol+TS & 1.62 & 2.15 & 1.84 & 0.65 & 2.13 & 1.99 & 2.81 & 0.35 
\\
BLYP & 1.59 & 3.07 & 1.80 & 0.08 & 2.27 & 1.41 & 2.92 & 0.66 
\\
BLYP+TS & 1.59 & 2.94 & 1.80 & 0.11 & 2.23 & 1.49 & 2.92 & 0.64 
\\
PW91 & 1.59 & 2.47 & 1.80 & 0.20 & 2.20 & 1.89 & 2.88 & 0.37 
\\
PW91+TS & 1.59 & 2.57 & 1.80 & 0.17 & 2.20 & 1.85 & 2.88 & 0.40 
\\
Experiment~\cite{soper_2013} & n/a & n/a & n/a & n/a & 2.43 & 1.34 & 2.97 & 0.69 
\\
\hline
 & \multicolumn{8}{c}{O-H RDF}
\\
PBE & 1.01 & 21.03 & 1.22 & 0.01 & 1.62 & 1.87 & 2.49 & 0.03 
\\
PBE+TS & 1.01 & 20.88 & 1.22 & 0.01 & 1.66 & 1.62 & 2.49 & 0.07 
\\
RPBE & 0.97 & 22.19 & 1.22 & 0.00 & 1.80 & 1.19 & 2.49 & 0.11 
\\
RPBE+TS & 0.97 & 21.20 & 1.22 & 0.00 & 1.84 & 1.67 & 2.41 & 0.07 
\\
revPBE & 0.97 & 21.44 & 1.26 & 0.00 & 1.73 & 1.32 & 2.52 & 0.09 
\\
revPBE+TS & 1.01 & 21.26 & 1.22 & 0.00 & 1.80 & 1.81 & 2.41 & 0.04 
\\
PBEsol & 1.01 & 18.00 & 1.26 & 0.33 & 1.55 & 2.18 & 2.49 & 0.03 
\\
PBEsol+TS & 1.01 & 18.23 & 1.26 & 0.29 & 1.55 & 2.12 & 2.41 & 0.03 
\\
BLYP & 0.97 & 24.51 & 1.22 & 0.00 & 1.77 & 1.12 & 2.45 & 0.17 
\\
BLYP+TS & 0.97 & 23.55 & 1.22 & 0.00 & 1.84 & 1.11 & 2.41 & 0.16 
\\
PW91 & 1.01 & 20.72 & 1.22 & 0.01 & 1.62 & 1.88 & 2.49 & 0.03 
\\
PW91+TS & 1.01 & 20.81 & 1.22 & 0.01 & 1.62 & 1.84 & 2.49 & 0.04 
\\
Experiment~\cite{soper_2013} & n/a & n/a & n/a & n/a & 1.86 & 1.04 & 2.46 & 0.21 
\\
\hline
 & \multicolumn{8}{c}{O-O RDF}
\\
PBE & 2.63 & 3.46 & 3.28 & 0.20 & 4.36 & 1.65 & 5.51 & 0.78 
\\
PBE+TS & 2.63 & 3.05 & 3.28 & 0.37 & 4.32 & 1.46 & 5.58 & 0.78 
\\
RPBE & 2.77 & 2.32 & 3.39 & 0.60 & 4.36 & 1.37 & 5.58 & 0.75 
\\
RPBE+TS & 2.77 & 3.18 & 3.28 & 0.43 & 4.43 & 1.34 & 5.48 & 0.81 
\\
revPBE & 2.70 & 2.56 & 3.35 & 0.49 & 4.40 & 1.45 & 5.69 & 0.74 
\\
revPBE+TS & 2.77 & 3.35 & 3.24 & 0.33 & 4.50 & 1.40 & 5.51 & 0.78 
\\
PBEsol & 2.56 & 4.13 & 3.24 & 0.18 & 4.29 & 1.64 & 5.55 & 0.75 
\\
PBEsol+TS & 2.56 & 4.07 & 3.24 & 0.20 & 4.29 & 1.59 & 5.22 & 0.83 
\\
BLYP & 2.74 & 2.20 & 3.42 & 0.73 & 4.36 & 1.22 & 5.58 & 0.85 
\\
BLYP+TS & 2.81 & 2.38 & 3.42 & 0.62 & 4.36 & 1.19 & 5.40 & 0.90 
\\
PW91 & 2.63 & 3.56 & 3.28 & 0.20 & 4.43 & 1.64 & 5.66 & 0.76 
\\
PW91+TS & 2.63 & 3.41 & 3.28 & 0.24 & 4.32 & 1.57 & 5.40 & 0.75 
\\
Experiment~\cite{soper_2013} & 2.79 & 2.50 & 3.39 & 0.80 & 4.53 & 1.12 & 5.58 & 0.88 
\end{tabular}
\end{ruledtabular}
\label{07}
\end{table*}

The extremely popular PBE benefits from dispersion corrections by improved description of all the studied properties. However, the
water overstructuring typical of the PBE functional cannot be completely suppressed with the inclusion of van der Waals corrections.
A popular strategy, often found in the literature to tackle this issue, is to perform MD with PBE at high temperature. The also
very popular BLYP functional offers by far the best description of water, in all the categories, in the absence of dispersion
corrections. When adding TS corrections the description is slightly worsened, in particular the value of the self-diffusion
coefficient. Unfortunately, BLYP is known to offer poor description of metallic systems~\cite{stroppa_2008}, which may limit its wider applicability.

All the functionals underestimate standard molar entropies by large factors, which vary between a best-case scenario of 17\% (BLYP)
and worst-case scenario of 44\% (PBEsol). Similarly, the self-diffusion coefficients are also underestimated by all the tested 
GGAs, varying between 10\% (BLYP) and 83\% (PBEsol). Unfortunately, it is likely that a fully satisfactory description of liquid water cannot
be achieved with any XC functional at this level of theory, although BLYP offers very good performance. With respect to dispersion corrections,
it seems that the description of water is worsened by adding them for those functionals which require largest corrections, i.e., functionals which
require small $s_R$ values: RPBE, revPBE and BLYP. The description of liquid water is improved by including dispersion corrections
for the other functionals: PBE, PW91 and PBEsol.

\section{Conclusion}

In conclusion, we have provided a list of optimal range parameters $s_R$ for different XC functionals based on minimization of the
MAE in the prediction of dispersion interactions of the S22 test set. These parameters allow computations of dispersion
corrections to DFT based on the Tkatchenko-Scheffler method for the following GGAs: PBE, RPBE, revPBE, PBEsol, BLYP, AM05 and PW91.
We have tested the effect of dispersion corrections on the description of liquid water offered by those functionals (except for AM05),
finding no systematic improvement, i.e., the improvement is strongly functional-dependent. Based on our simulations, we recommend
either BLYP for organic systems (due to its limitations to correctly describe metals), or RPBE more generally, as affordable options
to carry out atomistic studies of systems including water. When van der Waals interactions are expected
to play a significant role, for instance in the presence of solvated organic molecules and adsorption phenomena,
BLYP+TS and RPBE+TS can be good options, although the
improvement over other GGAs is not so obvious as for the uncorrected cases. PBEsol was observed to perform particularly badly for
liquid water, and we strongly advice against using PBEsol to simulate any system including an aqueous phase.

\begin{acknowledgments}
Computational resources for this project were provided by CSC--IT Center for Science through the Taito
supercluster. The author would like to thank Michael Walter for useful comments and email correspondence, as well
as being an author of the GPAW implementation of the TS scheme.
\end{acknowledgments}

\begin{widetext}
\appendix*

\section{\bf Code for $s_R$ optimization}

The following Python script can be used to optimize the $s_R$ parameter for a given XC functional (or list of functionals) with the
GPAW code:
\begin{python}
# This code was adapted by M.A. Caro from Michael Walter's TS documentation on the GPAW website
from __future__ import print_function
from ase import Atoms
from ase.parallel import paropen
from ase.data.s22 import data,s22
from ase.calculators.vdwcorrection import vdWTkatchenko09prl
from gpaw import GPAW, FermiDirac
from gpaw.cluster import Cluster
from gpaw.analyse.hirshfeld import HirshfeldPartitioning
from gpaw.analyse.vdwradii import vdWradii
import numpy as np

h = 0.18; box = 4.

sR_range = np.arange(0.88,1.02,0.02)
xc_list = ["PBE"]

for xc in xc_list:
    f = paropen('dispersion_energies_
    disp_en = {}
    for molecule in s22:
        disp_en[molecule] = []
        ss = Cluster(Atoms(data[molecule]['symbols'],
                     data[molecule]['positions']))
#       Split interacting system into the separate molecules
        s1 = ss.find_connected(0)
        s2 = ss.find_connected(-1)
        assert(len(ss) == len(s1) + len(s2))
        c = GPAW(xc=xc, h=h, nbands=-6, occupations=FermiDirac(width=0.1))
        for s in [s1, s2, ss]:
            s.set_calculator(c)
            s.minimal_box(box, h=h)
            s.get_potential_energy()
            E = {}
            for sR in sR_range:
                cc = vdWTkatchenko09prl(HirshfeldPartitioning(c),
                                        vdWradii(s.get_chemical_symbols(), xc))
                cc.sR = sR
                s.set_calculator(cc)
                E[sR] = s.get_potential_energy()
            disp_en[molecule].append(E)
#   Print
    for sR in sR_range:
        print("# sR = 
        for molecule in s22:
            ref = data[molecule]["interaction energy CC"]
            E1 = disp_en[molecule][0][sR]
            E2 = disp_en[molecule][1][sR]
            E12 = disp_en[molecule][2][sR]
#           Print molecule name, predicted interaction energy and reference CCSD(T) energy
            print("
        print("", file=f)
    f.close()
\end{python}
\end{widetext}

\end{document}